\documentstyle[twocolumn,epsf]{jpsj}

\def\runtitle{
Spin triplet superconductivity
 in Sr$_2$RuO$_4$
}

\def\runauthor{
Yasumasa {\sc Hasegawa}, Kazushige {\sc Machida} and Masa-aki {\sc Ozaki}
 }

\begin{document}
\title {
Spin triplet superconductivity with line nodes
 in Sr$_2$RuO$_4$
}

\author{
Yasumasa {\sc Hasegawa},  Kazushige {\sc Machida}$^{1}$ and Masa-aki {\sc
Ozaki}$^{2}$
}

\inst{
Faculty of Science, Himeji Institute of Technology,
Kamigori, Akou-gun, Hyogo 678-1297\newline
$^{1}$Department of Physics, Okayama University, Okayama 700-8530\newline
$^{2}$Department of Physics, Kochi University, Kochi 780-8072}

\recdate{September 22, 1999}

\abst{
Several possible odd-parity states are listed up group-theoretically and
examined in  light of recent experiments on Sr$_2$RuO$_4$.
Those include some of the $f$-wave pairing states, ${\mib d}({\mib
k})\propto{\hat{\mib z}}
  k_xk_y(k_x + {\rm i}k_y)$ and ${\hat{\mib z}}
  (k_x^2-k_y^2)(k_x + {\rm i}k_y)$ and other ${\hat{\mib z}}
  (k_x + {\rm i}k_y)\cos ck_z$ ($c$ is the $c$-axis lattice constant)
  as most plausible candidates. These are time-reversal symmetry broken
  states and have line nodes running either vertically (the former two) or
horizontally
  (the latter), consistent with experiments.
  Characterizations of these states and other possibilities are given.
}

\kword{
 spin triplet superconductivity, line node of the energy gap,
time-reversal symmetry breaking}

\sloppy
\maketitle


The possibility of the spin triplet pairing
 in Sr$_2$RuO$_4$ is pointed out\cite{rice95}
soon after the discovery of the superconductivity\cite{maeno94}.
Experimental evidence for the spin triplet pairing is accumulated.
The temperature-independent Knight shift for the external magnetic
field parallel to the conducting plane (the basal plane)
is thought to be the evidence for the spin triplet
pairing with the $d$-vector aligned to the $z$-axis\cite{ishida98}.
 The $\mu$SR experiment\cite{luke98}
indicates that the magnetic field is spontaneously induced
and the time reversal symmetry seems to be broken
in the superconducting state.
These
experiments were  explained by the $p$-wave spin triplet state with
the order parameter\cite{machida96,sigrist96},
\begin{equation}
  {\mib d}({\mib k})=\Delta_0 \hat{\mib z} (k_x + {\rm i}k_y),
 \label{pwave}
\end{equation}
where $\Delta_0$ is a constant. This state is analogous to the
Anderson-Brinkman-Morel (ABM) state, which is identified as the A
phase in superfluid $^3$He.\cite{leggett75} The AMB state in
superfluid $^3$He has point nodes of the energy gap at $k_x=k_y=0$ on the
spherical Fermi surface. However,  since  the Fermi surface of
Sr$_2$RuO$_4$ consists of three cylindrical
surfaces, the above $p$-wave state has a finite energy gap on the
Fermi surface.
The previous experiments of specific heat\cite{nishizaki98} $C(T)$
 and NMR relaxation
rate\cite{ishida97} $T_1^{-1}$
showed that about a half of the density of states remains at $T=0$
in the superconducting state. In order to explain this residual density
of states, the non-unitary states and the orbital dependent
superconductivity are proposed\cite{maeno96,agterberg97}.

However, it is reported in very recent
experiments
that the residual density of states are very small in the better
samples, i.e. $C(T) \propto T^2$ and $T_1^{-1} \propto T^3$ at low
temperatures\cite{ishidaprivate,maenoprivate}. These temperature
dependences are interpreted as a
consequence of the line nodes of the energy gap.
The residual density of states observed in the previous experiments
may be due to the impurity effects on the anisotropic
superconductivity with line nodes.

An anisotropic energy gap model caused by the finite range interaction
in the two-dimensional plane is proposed to explain the
line-node-like temperature dependences of the specific heat and
relaxation rate in Sr$_2$RuO$_4$\cite{miyake99}.
The order parameter in this case is
given by
\begin{equation}
  {\mib d}({\mib k})= \Delta_0 {\hat{\mib z}}
  (\sin a k_x + {\rm i} \sin a k_y) ,
 \label{pwave2}
\end{equation}
where $a$ is the lattice constant in the basal plane of the tetragonal
crystal D$_{4h}$.
This anisotropic energy gap  should give the
exponential temperature dependence as
$\exp (-\alpha \Delta /T)$  at temperatures below the smallest energy
gap $\Delta$  in clean samples,
where $\alpha$ is a constant of the order 1.
As far as we know, however,
 there are no experimental evidence for the existence of
the small gap in Sr$_2$RuO$_4$.

The previously classified pairing states\cite{machida96} with a line node
either (A) break the four-fold symmetry in the basal
plane such as ${\hat{\mib x}}k_x$ belonging to the two-dimensional
representation E$_u$, which is incompatible with the four-fold
symmetric H$_{c2}$ behavior\cite{mao99}, or (B) have the two component $d$
vector such as a non-unitary bipolar sate; ${\hat{\mib x}}k_x+i{\hat{\mib
y}}k_y$
which is incompatible with the Knight shift\cite{ishida98}
and $\mu$SR\cite{luke98} experiments.
Thus, if we accept these experiments, all the previous states becomes unlikely
to explain\cite{note}. This is also true for those states listed
by Rice and Sigrist\cite{rice95}
and Sigrist and Zhitomirsky\cite{sigrist96}.

Here we are going to first list up several possible odd-parity states
with lower angular momentum allowed under the tetragonal symmetry
D$_{4h}$ of Sr$_2$RuO$_4$, then we argue plausibility of these states in light
of the existing data where a pairing mechanism is also examined for
stabilizing
some of the classified states.

In Table I the orbital functions allowed in D$_{4h}$, which are
derived from the product of the two irreducible representations, are shown.
These belong to the two-dimensional representation E$_u$.
These are used when constructing the order parameters,
which are listed up in Table II.

\begin{table}
\caption{Orbital function in two-dimensional representation
E$_u$ and its product representation}
\begin{tabular}{cc}
\hline
Product&Orbital function: $(\lambda_1(k), \lambda_2(k))$\\
\hline
A$_{1g}$$\times$E$_u$&$(k_x^2+k_y^2)(k_x,k_y)$\\
A$_{2g}$$\times$E$_u$&$k_xk_y(k_x^2-k_y^2)(k_x,k_y)$\\
B$_{1g}$$\times$E$_u$&$(k_x^2-k_y^2)(k_x,k_y)$\\
B$_{2g}$$\times$E$_u$&$k_xk_y(k_x,k_y)$\\
\hline
\end{tabular}
(A$_{1g}$:1, $k_x^2+k_y^2$, A$_{2g}$: $k_xk_y(k_x^2-k_y^2)$,
B$_{1g}$: $k_x^2-k_y^2$,
B$_{2g}$: $k_xk_y$)
\label{table1}
\end{table}


\begin{table}
\caption{Order parameter and its characterization}
\begin{tabular}{ccc}
\hline
order parameter&Unitarity&TRSB\\
\hline
${\hat{\mib z}}\lambda_1(k)$&U&N\\
$({\hat{\mib x}}+i{\hat{\mib y}})\lambda_1(k)$&N&Y\\
$({\hat{\mib x}}+i{\hat{\mib y}})(\lambda_1(k)+\lambda_2(k))$&N&Y\\
${\hat{\mib z}}(\lambda_1(k)+\lambda_2(k))$&U&N\\
$({\hat{\mib x}}+i{\hat{\mib y}})(\lambda_1(k)+i\lambda_2(k))$&N&Y\\
${\hat{\mib x}}\lambda_1(k)+{\hat{\mib y}}\lambda_2(k)$&U&N\\
${\hat{\mib x}}\lambda_1(k)+i{\hat{\mib y}}\lambda_2(k)$&N&Y\\
\hline
\end{tabular}
Unitarity (U) or Non-unitary (N),
TRSB (time reversal symmetry breaking) yes (Y) or no (N)
\label{table2}
\end{table}


Among these allowed states the following $f$-pairing
states are particularly attractive:

\begin{eqnarray}
 {\mib d}({\mib k})&=&\Delta_0{\hat{\mib z}}(k^2_x-k^2_y)(k_x+ik_y)
\end{eqnarray}

\noindent
and

\begin{eqnarray}
 {\mib d}({\mib k})&=&\Delta_0{\hat{\mib z}}k_xk_y(k_x+ik_y)
\end{eqnarray}

\noindent
because
(1) they are both odd-parity states and have single $d$-vector
component, and (2) the four vertical line nodes run along the $c$-axis
and the four-fold symmetry in the basal plane is preserved.

Here we should point out a possibility that these $f$-pairing states
can mix $p$-wave pairing in general for D$_{4h}$, namely,

\begin{equation}
{\mib d}({\mib k})=\Delta_0 \hat{\mib z} 
 \{ k_xk_y(k_x + {\rm i}k_y) +\gamma (k_y +{\rm i}k_x) \}
\end{equation}

\noindent
with $\gamma$ being a complex number. This mixing washes out the desired
line node
except for some special case. Within the present framework we
assume that the angular decomposition is nearly complete, neglecting their
mixing. At the moment we do not know how well this decomposition is,
but the experimental facts seem to demand it.

So far we only consider the purely two-dimensional order parameters
characterized by the two wave vectors $k_x$ and $k_y$ in the basal plane.
However, there is another class of the
states compatible with the experimental data,
namely,

\begin{eqnarray}
  {\mib d}({\mib k})&=&{\hat {\mib z}}
   [f_o(k_x, k_y) (a_0 + a_1 \cos(c k_z) + a_2 \cos(2c k_z) + ... )
 \nonumber \\
    & & +  f_e(k_x, k_y) (b_1 \sin(c k_z) + b_2 \sin(2c k_z) + ... )],
\end{eqnarray}

\noindent
where the order parameter in this class
is not the partial wave in $k_z$ but the Fourier series
with $c$ being the lattice constant along the $c$-axis.
$f_o(k_x, k_y)$ and  $f_e(k_x, k_y)$ are the odd and even
functions with respect to the inversion
in the $k_x-k_y$ plane, respectively.
The previously considered states (eq.(\ref{pwave}) and eq.(\ref{pwave2})) are
two-dimensional states with $a_0=1$ and $a_n=b_n=0$ ($n >0$).
We propose the state with $a_1 \not= 0$ and other $a_n$ and $b_n$ are
zero. The order parameter is written as
\begin{equation}
 {\mib d}({\mib k}) =\Delta_0 {\hat {\mib z}} (k_x+i k_y)  \cos (c k_z).
\label{state1}
\end{equation}
This particular state can be regarded as derived from a product representation
A$_{1g}$$\times$E$_u$ and is periodic along the $c$-axis.
This lattice periodicity might be important because all the
three Fermi surfaces are open along the $c$-axis, but close within
the basal plane. In fact, this state has horizontal line nodes at $k_z=\pm
{\pi\over 2c}$,
which run parallel to the basal plane, and
breaks the time reversal symmetry.
Even if the $a_0$ term is mixed, i.e.
\begin{equation}
 {\mib d}({\mib k}) =\Delta_0 {\hat {\mib z}} (k_x+i k_y)  (\cos (c
k_z) + a_0),
\label{state11}
\end{equation}
the horizontal line nodes exist as long as $a_0$ is real and $|a_0|<1$.

If we take account of the tetragonal K$_2$NiF$_4$ type crystal
structure or the body centered tetragonal lattice of Ru,
the order parameter,
\begin{equation}
 {\mib d}({\mib k}) = \Delta_0 {\hat {\mib z}}
  (\sin \frac{a k_x}{2}+{\rm i} \sin \frac{a k_y}{2})
  \cos \frac{c k_z}{2}  ,
\label{state2}
\end{equation}
is preferable.
This state has the horizontal line
nodes at  $k_z=\pm {\pi\over c}$.
This state is possible if the effective
intralayer interaction is repulsive and interlayer coupling is
attractive, as we discuss below.

Due to the Coulomb interaction
the effective interaction will be repulsive  between electrons
in the same plane.
We assume the effective
interaction is attractive for electrons at $r_i$ and $r_i \pm {a\over 2} \pm
{a\over 2}\pm {c\over 2}$,
\begin{equation}
  V({\mib r},{\mib r}') = -V_0, \hspace{0.5cm}
 \mbox{ if ${\mib r}'={\mib r}\pm
\frac{a}{2}\hat{\mib x}
  \pm \frac{a}{2} \hat{\mib y} \pm \frac{c}{2} \hat{\mib z}$}
\end{equation}
where $V_0 >0$, $r_i$ and $r_i \pm {a\over 2}\hat{\mib x}  \pm
{a\over 2} \hat{\mib y} \pm {c\over 2}\hat{\mib z}$ are the lattice sites
of Ru, and $a$ and $c$ are the lattice constant in the body centered
tetragonal lattice.
The Fourier transform of the effective interaction is
\begin{eqnarray}
  V({\mib k},{\mib k}')&=& -8V_0
     \cos \frac{a(k_x-k_x')}{2}
     \cos \frac{a(k_y-k_y')}{2}
\nonumber \\ & & \times
     \cos \frac{c(k_z-k_z')}{2}
 \nonumber \\
  &=&
  -8V_0(\cos\frac{ak_x}{2} \cos\frac{ak_x'}{2}
       +\sin\frac{ak_x}{2} \sin\frac{ak_x'}{2})
 \nonumber \\
  & & \times
       (\cos\frac{ak_y}{2} \cos\frac{ak_y'}{2}
       +\sin\frac{ak_y}{2} \sin\frac{ak_y'}{2})
 \nonumber \\
  & & \times
       (\cos\frac{ck_z}{2} \cos\frac{ck_z'}{2}
       +\sin\frac{ck_z}{2} \sin\frac{ck_z'}{2}) .
\end{eqnarray}
The orbital part of
the order parameter caused by this interaction
should have the form
\begin{equation}
  \left \{
   \begin{array}{l}
    \sin\frac{ak_x}{2}\cos\frac{ak_y}{2}\cos\frac{ck_z}{2} \\
    \cos\frac{ak_x}{2}\sin\frac{ak_y}{2}\cos\frac{ck_z}{2} \\
    \cos\frac{ak_x}{2}\cos\frac{ak_y}{2}\sin\frac{ck_z}{2} \\
    \sin\frac{ak_x}{2}\sin\frac{ak_y}{2}\sin\frac{ck_z}{2}
   \end{array} \right. ,
\end{equation}
in the triplet pairing.
These four states have different transition temperature in the weak
coupling limit, but
the first two states are degenerate.
Although  we have to calculate in more detail to find which state
is most stable,  we think the
order parameter
 \begin{eqnarray}
  {\mib d}({\mib k}) &=& \hat{\mib z} \Delta_0
  (  \sin \frac{ak_x}{2} \cos\frac{ak_y}{2}
  +{\rm i} \cos \frac{ak_x}{2} \sin\frac{ak_y}{2})
  \nonumber \\
 & & \times
 \cos\frac{ck_z}{2}
 \label{state3}
\end{eqnarray}
will possibly be the stable state. If we neglect the $k_x$ and $k_y$
dependences in $\cos \frac{a k_x}{2}$
and  $\cos \frac{a k_y}{2}$, we get the order parameter given
in eq.(\ref{state2}).

We propose some experiments to test the horizontal and vertical line nodes.
The thermal conductivity will have weak dependence on the direction
in the $x-y$ plane and have a power low dependence on temperature,
if the line nodes are horizontal. On the other hand  it will
show strong four-fold symmetry and exponential dependences in some
direction if the superconducting state has the vertical nodes of the
energy gap. The penetration depth will also show the orientation of
the line node. This kind of the orientation dependent transport measurements,
including the ultrasound attenuation was decisive in identifying
the intricate gap structure of another triplet superconductor
UPt$_3$\cite{nishira}.

The Josephson effect between conventional superconductors and
Sr$_2$RuO$_4$ is observed and shows the interesting
temperature dependences\cite{jin99}.
Some explanations\cite{honerkamp98,yamashiro98} are given.
Although the experimental study of the
direction-dependence of the Josephson effect will be difficult because
of the roughness of the interface, the existence of the Ru lamellas
and 3K phases,\cite{sumiyama00} we consider the ideal Josephson junction.
Since all the states considered in this paper have ${\mib d}
\parallel {\hat{\mib z}}$, the Josephson current $I \parallel
\hat{\mib z}$ between the triplet and the singlet
superconductors is forbidden in the first order in the spin-orbit
coupling.\cite{geshkenbein86,hasegawa98} The Josephson current
between spin triplet and spin singlet is possible,
if the ${\mib d}$ vector is not
parallel to the current and the total (spin plus angular) momentum is
conserved. In the states given in eq.(\ref{state1}), eq.(\ref{state2})
or eq.(\ref{state3}), the Josephson current in the $x-y$ plane
is canceled, since the order parameter changes the sign in the
$k_z$-direction. However, the Fermi surface is corrugated as observed by
angle-dependent
magnetoresistance oscillation\cite{yoshida98b,ohmichi99} and
the angle dependence of the de Haas van Alphen effect\cite{yoshida98},
and as a result
the cancellation will not be  perfect.
In the presence of the
 mixing of the $k_z$
independent term (as in eq.(\ref{state11})) the Josephson current is
also finite.

The horizontal line nodes will be most clearly seen by the neutron scattering.
Since the quasi-particles are excited around $k_z=\pm {\pi\over c}$ in the
superconducting state eq.(\ref{state2}) or eq.(\ref{state3}),
the excitations between these quasi-particles,
i.e. excitations with $q_z=\pm{2\pi\over c}$ will become large. The neutron
scattering experiment below $T_c$ will have peaks at
$q_z=\pm {2\pi\over c}$, while it is almost independent of $q_z$ above $T_c$.

As a final remark, we point out the following additional possibilities:

(1) Suppose that we disregard the $\mu$SR experiment\cite{luke98}.
Among the previous states\cite{machida96} the non-unitary bipolar state,
described by

\begin{equation}
  {\mib d}({\mib k}) =\Delta_0(\hat{\mib x}k_x + {\rm i}\hat{\mib y}k_y)
\end{equation}

\noindent
 has line nodes for both up-spin and down-spin pair
branches, thus there is no
residual $T$-linear specific heat.
 This state
does not exhibit the macroscopic spontaneous moment
averaged over the Fermi surfaces below $T_c$.
Since the effective spin-orbit coupling felt by the Cooper pairs may
be small in Sr$_2$RuO$_4$\cite{machida96}, the $d$ vector can rotate
in the spin space. When the external field $\mib H$ is in the basal
plane, the non-unitary bipolar state becomes

\begin{equation}
  {\mib d}({\mib k}) =\Delta_0(\hat{\mib z}k_x + {\rm i}\hat{\mib j}k_y)
\end{equation}

\noindent
with $\hat{\mib j}$ being a vector lying in the basal plane which
is rotatable so as to keep ${\mib H}\perp \hat{\mib j}$,
being consistent with the Knight shift
experiment\cite{ishida98}.
As another possibility, the time reversal symmetry conserved state becomes a
candidate
such as $\hat{\mib z}k_x$ or $\hat{\mib z}k_y$ which are degenerate in
D$_{4h}$.
They form a domain structure in real system which may preserve the overall
four-fold symmetry. Note that there is no four-fold symmetric states without
the variable $k_z$ in our classified states\cite{machida96}.

(2) Suppose further that we disregard the Knight shift
experiment\cite{ishida98}
in addition to the
 $\mu$SR experiment\cite{luke98}. Then the following singlet pairings with
line nodes are
 another candidates, including

 \begin{equation}
\Delta(k) =\Delta_0k_xk_y(k_x^2-k_y^2) \\ \ \ \ \ {\rm  for \\ \  A_{2g}}
\end{equation}

 \begin{equation}
\Delta(k) =\Delta_0(k_x^2-k_y^2) \\ \ \ \ \ {\rm for \\\ B_{1g}}
\end{equation}

 \begin{equation}
\Delta(k) =\Delta_0k_xk_y \\ \ \ \ \ {\rm for \\\ B_{2g}}.
\end{equation}

\noindent
These singlet states might be stabilized\cite{mazin99} by the observed
antiferromagnetic fluctuations\cite{sidis99} whose wave vector is
${\bf Q}\sim ({2\pi\over 3a},{2\pi\over 3a},0)$.

In conclusion, we list up and examine
the several possible superconducting states in
Sr$_2$RuO$_4$, including
some of the $f$-wave pairing states, ${\hat{\mib z}}
  k_xk_y(k_x + {\rm i}k_y)$ and ${\hat{\mib z}}
  (k_x^2-k_y^2)(k_x + {\rm i}k_y)$ and other ${\hat{\mib z}}
  (k_x + {\rm i}k_y)\cos ck_z$.
  These states explain the experiments of  $\mu$SR, NMR
and the specific heat.
Experiments to distinguish these states are proposed.

The authors would like to thank  Y. Maeno, K. Ishida,
A. Sumiyama, M. Ichioka and A.G. Lebed for helpful discussions.



\begin{thebibliography}{99}
\bibitem{note}
Note that we cannot improve this situation by extending the
$p$-wave basis function to the periodic form: $k_x\rightarrow\sin ak_x$
$k_y\rightarrow\sin ak_y$ such as shown in eq.(\ref{pwave2})
\end{thebibliography}
\end{document}